\documentclass[a4paper]{article}

\usepackage[utf8]{inputenc}

\usepackage{multicol}
\usepackage{amsthm}
\usepackage{bm}
\usepackage{physics}

\usepackage[dvipdfmx]{graphicx}

\usepackage{bmpsize}
\usepackage{amssymb}
\usepackage{amsmath}
\usepackage{color}
\usepackage{amsthm}
\usepackage{comment}

\newcommand{\BA}{\begin{eqnarray}}
\newcommand{\EA}{\end{eqnarray}}

\definecolor{dgreen}{rgb}{0.0, 0.5, 0.0}

\begin{document}

\fontsize{14pt}{16.5pt}\selectfont

\begin{center}
\bf{
Emergence of ultradiscrete states due to phase lock\\ 
caused by saddle-node bifurcation in discrete limit cycles
}
\end{center}
\fontsize{12pt}{11pt}\selectfont
\begin{center}
Yoshihiro Yamazaki$^{1 *)}$ and Shousuke Ohmori$^{2, 3}$\\ 
\end{center}

\vspace{1mm}

\noindent
$^1$\it{Department of Physics, Waseda University, Shinjuku, Tokyo 169-8555, Japan}\\
$^2$\it{National Institute of Technology, Gunma College, Maebashi-shi, Gunma 371-8530, Japan}\\
$^3$\it{Waseda Research Institute for Science and Engineering, Waseda University}{Shinjuku, Tokyo 169-8555, Japan}\\

\noindent
*corresponding author : yoshy@waseda.jp\\
~~\\
\rm
\fontsize{11pt}{14pt}\selectfont\noindent

\baselineskip 30pt

\noindent
{\bf Abstract}\\
%
%
	%
	Dynamical properties of limit cycles in a tropically discretized 
	negative feedback model are numerically investigated.
	This model has a controlling parameter $\tau$, 
	which corresponds to time interval for the time evolution 
	of phase in the limit cycles.
	By considering $\tau$ as a bifurcation parameter, 
	we find that ultradiscrete state emerges due to phase lock 
	caused by saddle-node bifurcation.
	Furthermore, focusing on limit cycles  
	for the max-plus negative feedback model, 
	it is found that the unstable limit cycle in the max-plus model corresponds to 
	the unstable fixed points emerging  
	by the saddle-node bifurcation 
	in the tropically discretized model.

\bigskip

\bigskip




%
%

\textit{Introduction}: 
Limit cycles can be encountered in various systems 
as one of fundamental nonlinear phenomena\cite{Strogatz}, 
and various models have been proposed to reproduce limit cycles such as 
predator-prey model in ecological systems, 
negative feedback model and Fitzhugh-Nagumo model in biological systems, 
Sel'kov model in biochemical reactions, 
van der Pol equation in electric circuits.
%
These models are mainly expressed 
in terms of continuous differential or discrete difference equations 
according to the time evolution of respective real phenomena.
Furthermore, there also have been studies to derive max-plus equations 
for the above models 
by appropriate discretization and ultradiscretization, 
so that derived max-plus equations can be interpreted 
as cellular automata
\cite{Willox2007,Gibo2015,Ohmori2016,Ohmori2021,Yamazaki2021,
Ohmori2022,Ohmori2023c,Isojima2022}.
The interesting point here is that 
derived max-plus equations can also possess  
similar limit cycle solutions 
to the continuous or discretized equations.
Thus, it is important to clarify 
how the limit cycle structures are caused and are retained 
and how the limit cycles are ultradiscretized 
in discretized and max-plus systems.

%

%
As an appropriate discretization, 
tropical discretization is typically adopted.
Here we briefly introduce the tropical discretization\cite{Murata2013}, 
which is a discretizing procedure 
converting a continuous differential equation 
into a discrete difference equation with only positive variables.
Now we focus on a two-variable dynamical system having limit cycle solutions, 
and the following type of equations is treated  
for positive values of $x$ and $y$,  
\begin{equation}
	\frac{dx}{dt} = f_{1}(x, y)-g_{1}(x, y), \;\;\;\;\;  
	\frac{dy}{dt} = f_{2}(x, y)-g_{2}(x, y), 
	\label{eqn:1-1} 
\end{equation} 
where $f_{j}$ and $g_{j}$ are positive functions ($j = 1, 2$).
The tropical discretization of eq. (\ref{eqn:1-1}) is 
given as 
\begin{equation}
	x _{n+1}=x _n \frac{x _n+\tau f_{1}(x _n, y_n)}{x _n+\tau g_{1}(x _n, y_n)}, \;\;\;\;\; 
	y _{n+1}=y _n \frac{y _n+\tau f_{2}(x _n, y_n)}{y _n+\tau g_{2}(x _n, y_n)},
	\label{eqn:1-2}
\end{equation}
where $x_{n} = x(n\tau)$, $y_{n} = y(n\tau)$, and 
$\tau ( > 0 ) $ and $n$ show the discretized time interval and 
the number of iteration steps, respectively.
Note that eq. (\ref{eqn:1-1}) can be reproduced 
from eq. (\ref{eqn:1-2})  by taking $\tau \rightarrow0$. 
%

%
%
For application of the tropical discretization 
to dynamical systems with limit cycle solutions, 
Carstea et al. treated the three dimensional model for 
a reaction of an organism to pathogen invasion 
and inflammatory response\cite{Carstea2006}.
(They called it PRE model.
P, R, and E stand for the initial letters of pathogens, responders, and effectors, respectively.)
They reported that PRE model possesses limit cycle solutions 
and that the ultradiscrete max-plus equations 
obtained from the PRE model also have limit cycle solutions.
Based on this study, Willox et al. further investigated 
dynamical properties of the limit cycles 
by the ultradiscretized PRE model\cite{Willox2007}.
They also investigated a two dimensional predator-prey model 
as a more simplified one by adopting Fourier spectrum analysis.
They suggested that ultradiscrete limit cycles correspond to 
a limit case of the tropically discretized ones.
%

Gibo and Ito studied a negative feedback model\cite{Gibo2015}.
For the continuous negative feedback model, 
it has been confirmed that 
there is no limit cycle solution\cite{Griffith1968}.
However they showed that the tropically discretized 
negative feedback model exhibits Neimark-Sacker bifurcation 
and has limit cycle solutions\cite{Gibo2015}.
They pointed out that in some biological systems, 
the discrete model with limit cycles is better suited 
to real phenomena.
Furthermore, they derived the max-plus negative feedback model 
and numerically showed the existence of the attractive ultradiscre limit cycle,  
which consists of four states.
Based on the study of Gibo and Ito, 
we further investigated the dynamical properties of these 
negative feedback models\cite{Ohmori2023c} 
and showed emergence of the ultradiscre limit cycle for large $\tau$.
And by using the Poincar\'{e} map method, we found that 
the max-plus model has the two limit cycles, 
stable (attractive) and unstable (repulsive).

We have also investigated dynamical properties of 
limit cycles obtained from Sel'kov model
\cite{Selkov1968,Ohmori2021,Yamazaki2021,Ohmori2022}.
We confirmed that ultradiscrete states emerge for large $\tau$ 
in the tropically discretized Sel'kov model.
Also we found that there exist not only a stable limit cycle 
but also an unstable one  
in the max-plus equations obtained via ultradiscretization 
in the limit of $\tau \to \infty$.
These properties are essentially the same as those 
of the negative feedback models shown above.

In view of these previous studies, 
it is important to advance analysis for the dynamical properties 
of the limit cycles in these discrete models.
Actually, the previous analyses seem to be insufficient 
especially for understanding emergence of ultradiscrete states 
in tropically discretized systems.
And relationship between limit cycle states obtained from 
tropically discretized and max-plus models is not clear.
Furthermore, there is no explanation for existence 
of the unstable limit cycle solutions in the max-plus models.
In this letter, focusing on limit cycle solutions 
of the tropically discretized negative feedback model,
we show one scenario of how ultradiscretized states emerge  
in the limit cycles with $\tau$ as a bifurcation parameter 
from the viewpoint of bifurcation phenomena.
And we discuss the correspondence with the limit cycles 
obtained from the max-plus negative feedback model.

%
\begin{figure}[b!]
	\begin{center}
		\includegraphics[width=6.5cm]{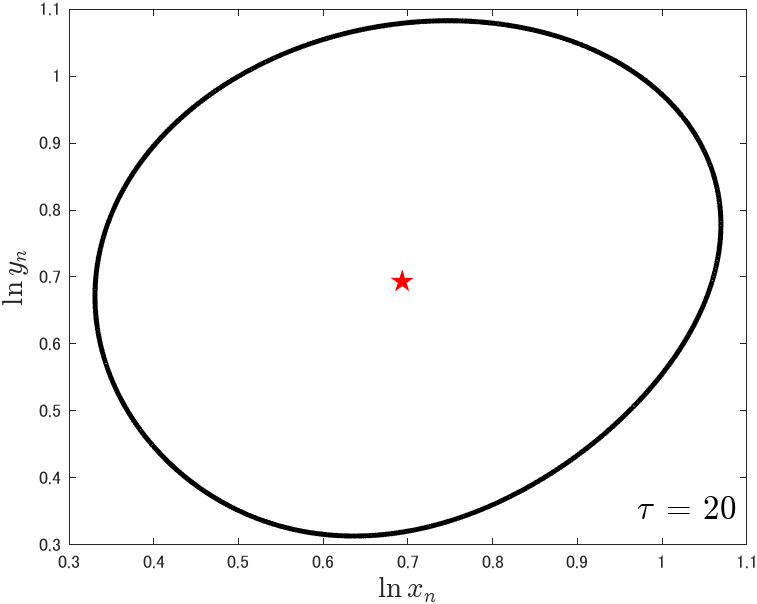}
		\includegraphics[width=6.5cm]{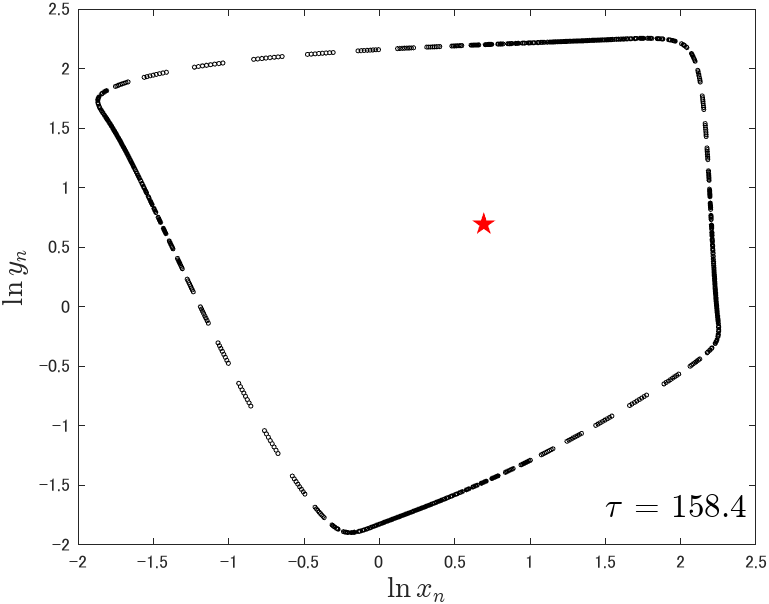}
	\end{center}
	\vspace{-8pt}
	\hspace{4.2cm}
	(a)
	\hspace{6.5cm}
	(b)	
	\begin{center}
		\includegraphics[width=6.5cm]{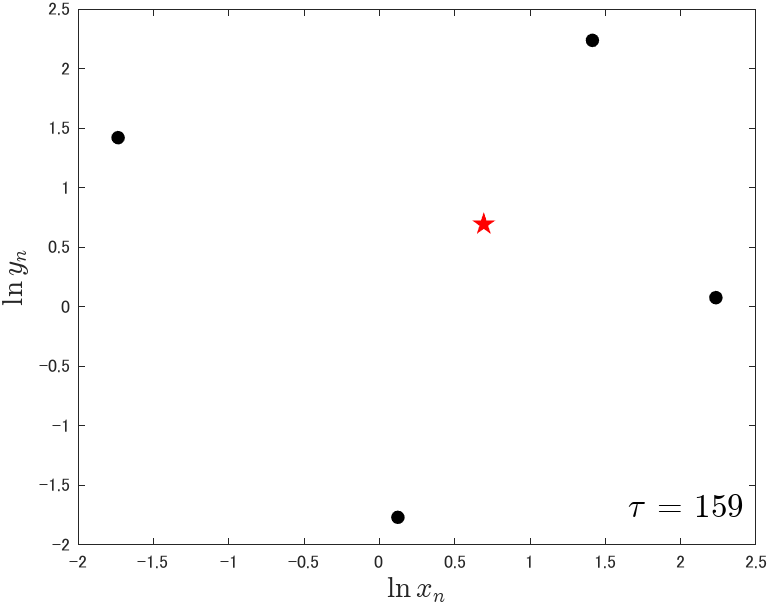}
		\includegraphics[width=6.5cm]{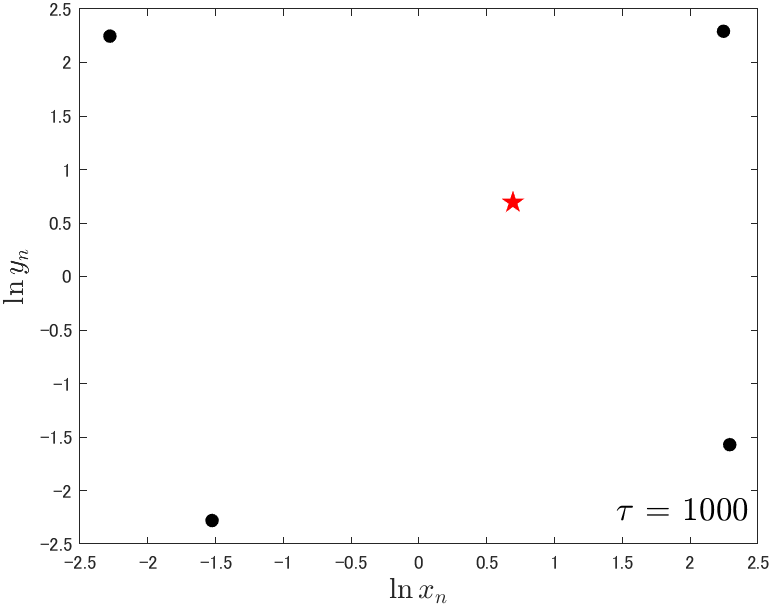}
	\end{center}
	\vspace{-8pt}
	\hspace{4.2cm}
	(c)
	\hspace{6.5cm}
	(d)	
	\caption{\label{fig:limit_cycles} 
	  The limit cycles obtained from eq.(\ref{eqn:tdnfb-x})
		with $b=10$ and $m=2$.
		The values of $\tau$ are (a) 20, (b) 158.4, 
		(c) 159, and (d) 1000. 
		In (c) and (d), the limit cycles consist of the four states, 
		which are depicted by the black filled cirlces.
		These are plotted with log-log scales.
		The red star in each figure shows the fixed point $(\bar x, \bar y)=(2, 2)$.}
	\end{figure}
%

\textit{Modelling and numerical results}: 
Let us start with introducing 
the tropically discretized negative feedback model 
for $(x_{n}, y_{n})$\cite{Gibo2015,Ohmori2023c}: 
\begin{equation}
	x_{n+1} = \frac{x_n+\tau y_n}{1+\tau} \equiv \eta_{\tau}(x _n, y_n), 
	\;\;\;\;
	y_{n+1} = \frac{y_n + \frac{\tau}{1+x_n^m}}{1+ \frac{\tau}{b}}
		\equiv \xi_{\tau}(x _n, y_n), 
	\label{eqn:tdnfb-x}
\end{equation}
where $b$ and $m$ are positive parameters.
We set $m=2$ and $b=10$ hereafter for numerical calculation, 
which was done by using Mathematica and MATLAB. 
Note that the original continuous equations for eq.(\ref{eqn:tdnfb-x}), 
$dx/dt = y - x$ and $dy/dt = 1/(1+x^{m}) - y/b$, possess 
the unique stable fixed point and 
there is no limit cycle solution\cite{Griffith1968}.
On the other hand, following the previous results 
in refs.\cite{Gibo2015,Ohmori2023c,Ohmori2023a}, 
eq.(\ref{eqn:tdnfb-x}) exhibits Neimark-Sacker bifurcation 
at $\tau = \tau_{0} = \frac{b(b+1)}{(m-1)b-m\bar{x}} = \frac{55}{3}$, 
and has a limit cycle solution for $\tau > \tau_{0}$ 
even in the limit of $\tau \to +\infty$.
Figure \ref{fig:limit_cycles} shows examples of the limit cycles 
with four different values of $\tau$.
It is found that as $\tau$ increases phase states 
in the limit cycles tend to be sparsely distributed and 
finally converge to the four states, 
which correspond to the ultradiscrete states. 
These results are consistent with our previous study\cite{Ohmori2023c}.

%
\begin{figure}[h!]
	\begin{center}
		\includegraphics[width=7cm]{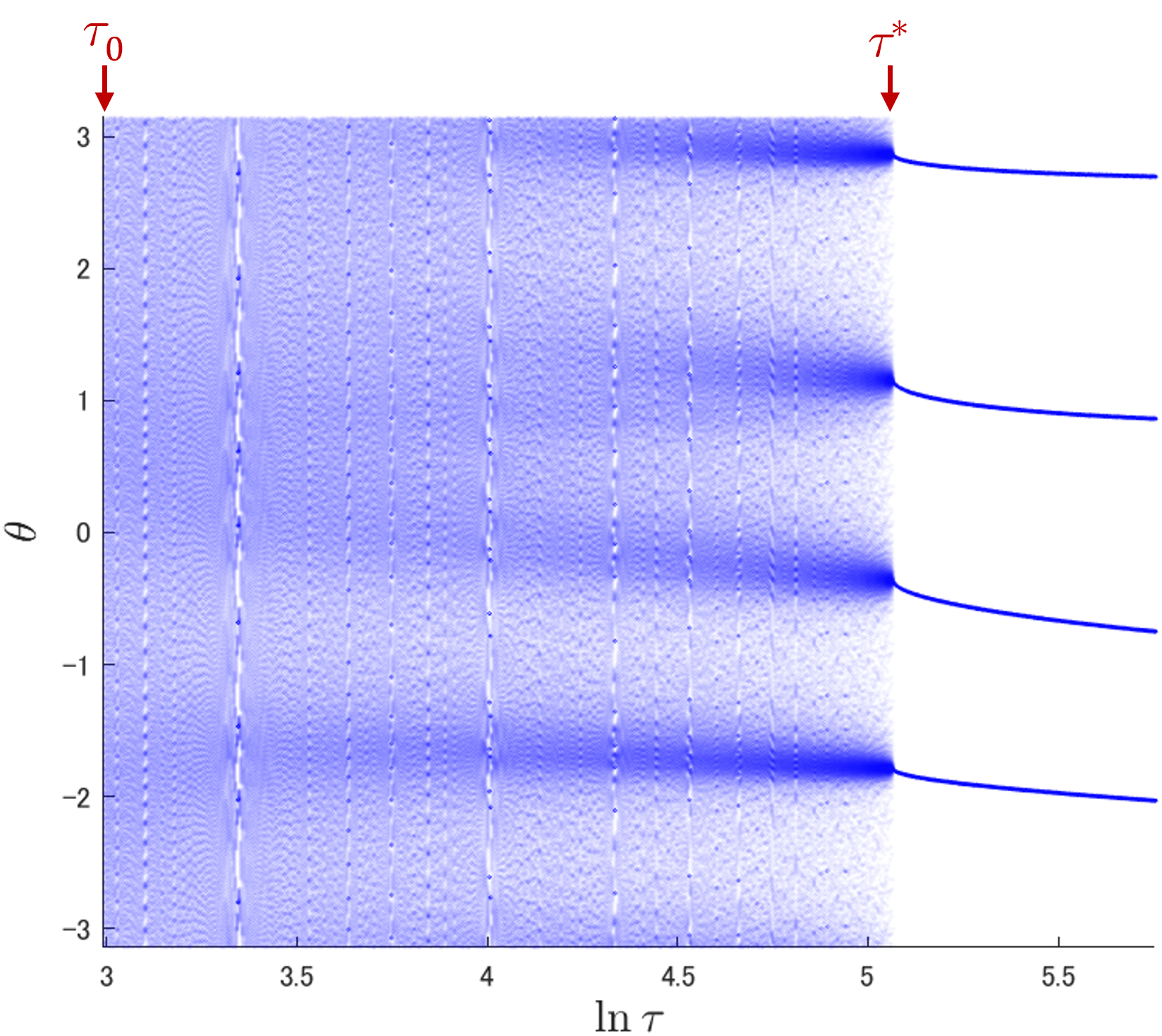}
		\includegraphics[width=7cm]{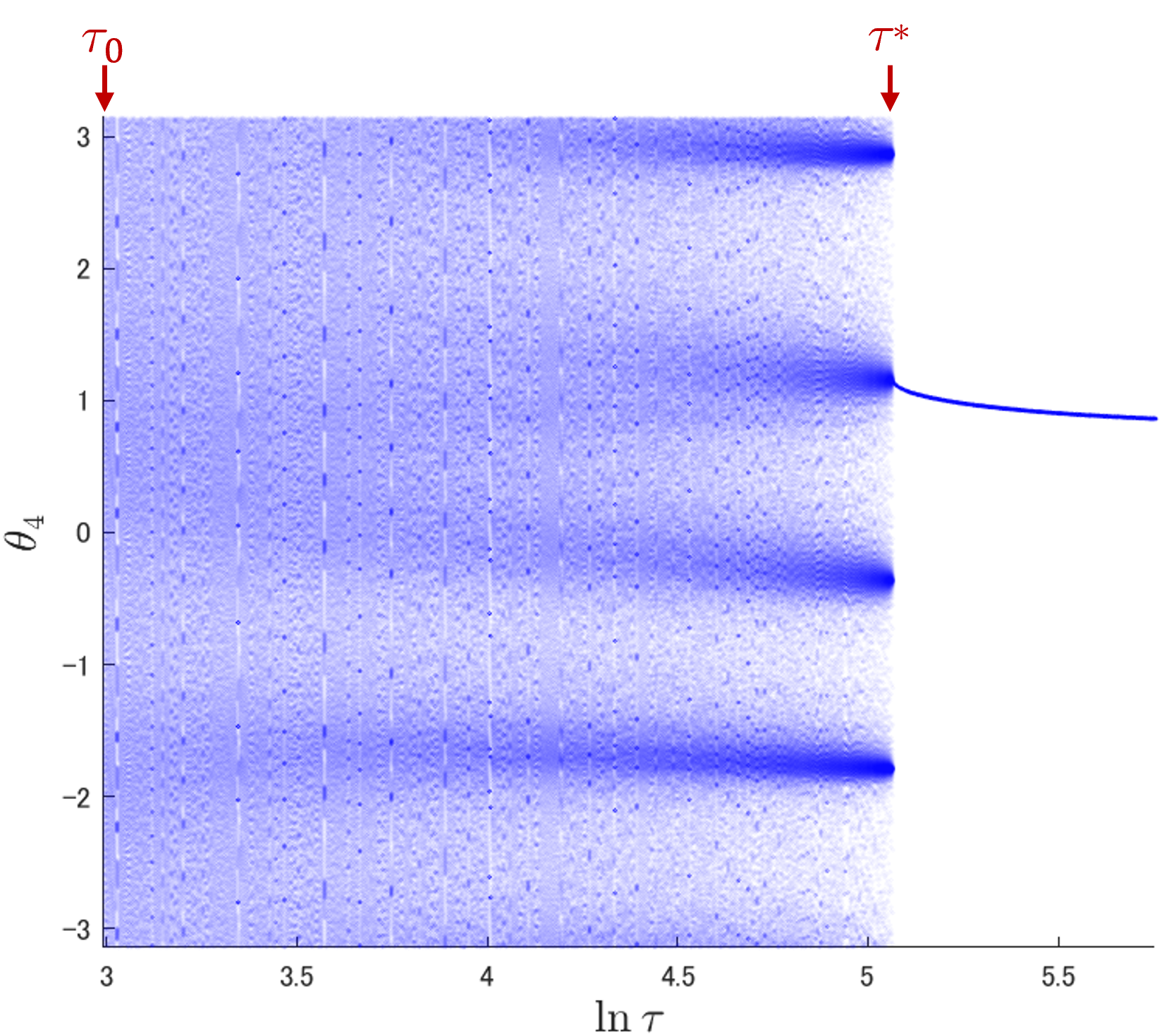}
	\end{center}
	\vspace{-8pt}
	\hspace{4.2cm}
	(a)
	\hspace{6.5cm}
	(b)	

	\caption{\label{fig:lc_deltat} 
	  (a) The scatter plot of the phase $\{\theta_{n}(\tau)\}$ 
		in the limit cycles, where $m=2$ and $b=10$. 
		$\theta_{n}(\tau)$ is defined as eq.(\ref{eqn:def_theta}). 
		(b) The scatter plot of the phase $\{\theta_{4n}(\tau)\}$.
		In both figures, 500 points are plotted at each value of $\tau$.}
\end{figure}

\begin{figure}[t!]
	\begin{center}
		\includegraphics[width=7cm]{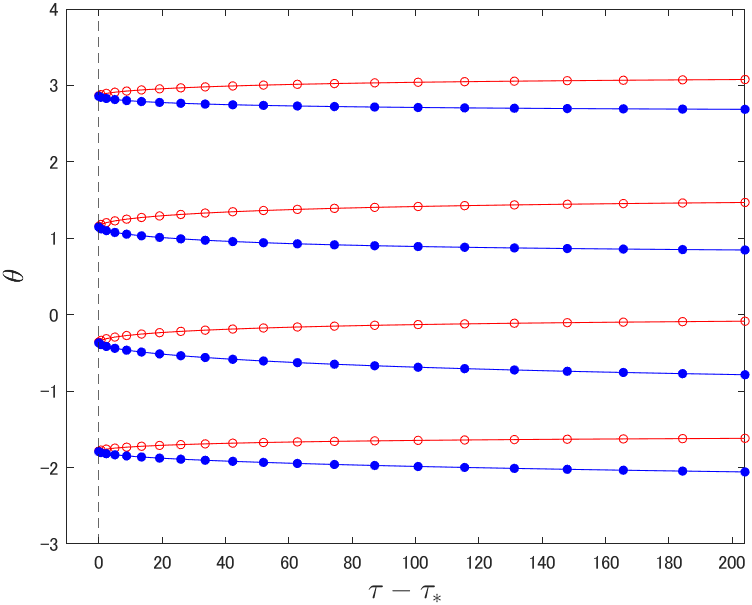}
	\end{center}
	\caption{\label{fig:Phase_st_unst} 
		Phase description of the eight fixed points
		$\{\bar{\theta}_{4}^{\rm{(s)}}\}$ (blue filled circles) and 
		$\{\bar{\theta}_{4}^{\rm{(u)}}\}$ (red open circles) 
		as a function of $\tau (> \tau^{\ast})$.}
\end{figure}

For the state $(x_{n}, y_{n})$ in the limit cycle, 
we define the phase $\theta_{n}(\tau)$ as
\begin{equation}
	\theta_{n}(\tau) = \arctan 
	\displaystyle \frac{\ln y_{n}-\ln \bar{y}}{\ln x_{n}-\ln \bar{x}}, 
	\label{eqn:def_theta}
\end{equation}
which takes a value in the range of $[-\pi, +\pi]$.
Figure \ref{fig:lc_deltat}(a) shows the scatter plot 
of $\{ \theta_{n}(\tau) \}$ in the limit cycles as a function of $\tau$.
It is clearly found that distribution of $\{ \theta_{n}(\tau) \}$ 
drastically changes at a certain value of $\tau$, 
denoted by $\tau^{\ast}$ in Fig. \ref{fig:lc_deltat}.
Note that Fig. \ref{fig:lc_deltat}(a) can be considered as 
a bifurcation diagram of $\{ \theta_{n}(\tau) \}$ 
for eq.(\ref{eqn:tdnfb-x}), and 
the change at $\tau = \tau^{\ast}$ suggests transition between 
discrete and ultradiscrete limit cycles.
Since the ultradiscrete limit cycle consists of four states, 
we focus on the phase state distribution of every four steps,  
$\{ \theta_{4n}\} (n = 1, 2, 3, \cdots)$ as shown in 
Fig.\ref{fig:lc_deltat}(b).
$\{ \theta_{4n} \}$ is obtained from 
4-th iterates of $\eta_{\tau}$ and $\xi_{\tau}$, 
denoted by $\eta^{4}_{\tau}$ and $\xi^{4}_{\tau}$.
In Fig. \ref{fig:lc_deltat}(b), 
the value of $\{ \theta_{4n}\}$ for $\tau > \tau^{\ast}$
corresponds to one of the stable fixed points for 
\begin{equation}
	x_{4(n+1)}  =  \eta^{4}_{\tau}(x_{4n}, y_{4n}), 
	\;\;\;\;
	y_{4(n+1)}  =  \xi^{4}_{\tau}(x_{4n}, y_{4n}). 
	\label{eqn:tdnfb-x-4}
\end{equation}
Actually except for $(\bar{x}, \bar{y})$, 
we find eight fixed points 
$(\bar{x}_{4}, \bar{y}_{4})$, which are obtained from 
$\bar{x}_{4} = \eta^{4}_{\tau}(\bar{x}_{4}, \bar{y}_{4})$ 
and $\bar{y}_{4} = \xi^{4}_{\tau}(\bar{x}_{4}, \bar{y}_{4})$ 
when $\tau > \tau^{\ast}$, whereas 
there is no fixed point for $\tau < \tau^{\ast}$.
The value of $\tau^{\ast}$ was numerically estimated 
as $\tau^{\ast} \approx 158.7137871989345$ 
from the change in the number of the fixed points.
Furthermore, it is found that four of the eight fixed points 
coincide with the four lines shown in Fig.\ref{fig:lc_deltat}(a) 
for $\tau > \tau^{\ast}$.
Now we denote these four points as 
$\{\bar{\theta}_{4}^{\rm{(s)}}(\tau)\}$, and 
the rest four are denoted 
as $\{\bar{\theta}_{4}^{\rm{(u)}}(\tau)\}$, 
which are identified later.
Figure \ref{fig:Phase_st_unst} shows the fixed points 
$\{\bar{\theta}_{4}^{\rm{(s)}}\}$ (blue filled circles) and 
$\{\bar{\theta}_{4}^{\rm{(u)}}\}$ (red open circles) 
as a function of $\tau$.
It is found from this figure that each value in $\{\bar{\theta}_{4}^{\rm{(s)}}\}$ is 
paired with one value in $\{\bar{\theta}_{4}^{\rm{(u)}}\}$, 
and they coalesce and vanish at $\tau = \tau^{\ast}$.

\begin{figure}[t!]
	\begin{center}
		\includegraphics[width=7cm]{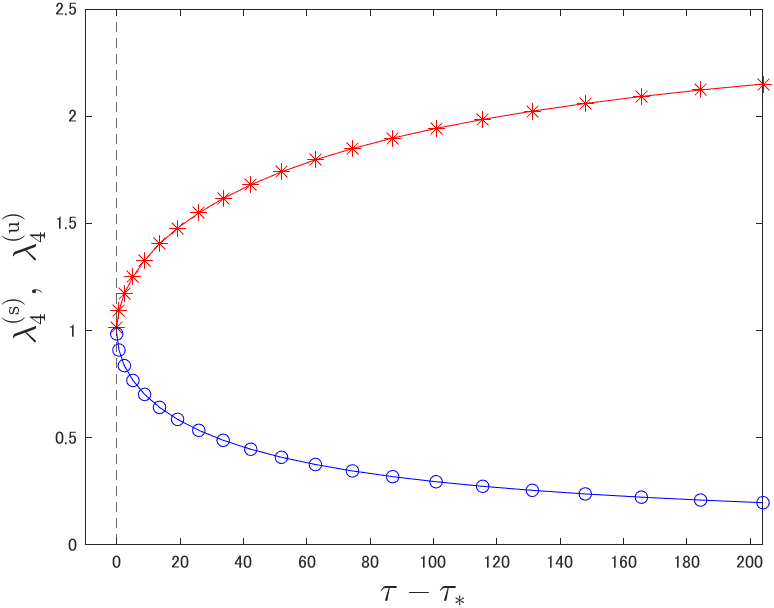}
		\hspace{0.5cm}
		\includegraphics[width=7cm]{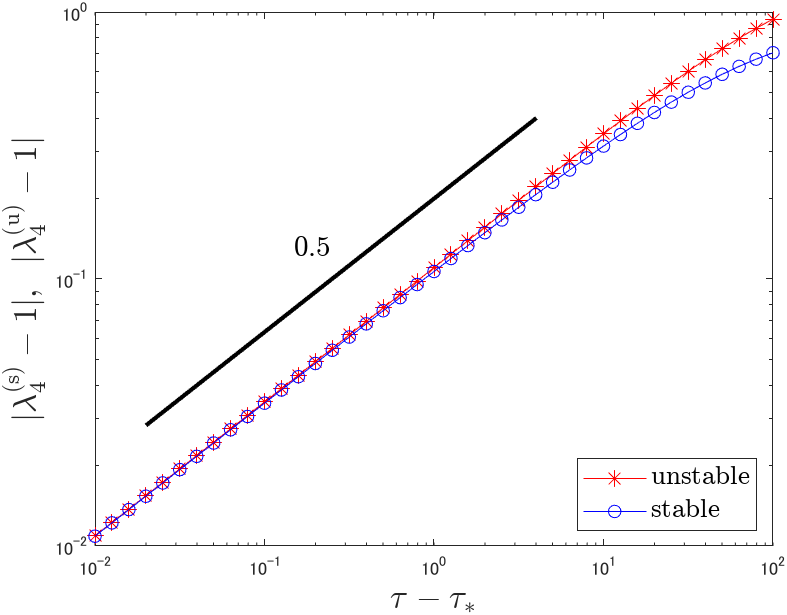}
	\end{center}
	\vspace{-8pt}
	\hspace{4.2cm}
	(a)
	\hspace{6.5cm}
	(b)	

	\caption{\label{fig:Lambda_st_unst} 
	  (a) The maximum eigenvalues $\lambda_{4}^{\rm{(s)}}$ for 
		$\{\bar{\theta}_{4}^{\rm{(s)}}\}$ and $\lambda_{4}^{\rm{(u)}}$ for 
		$\{\bar{\theta}_{4}^{\rm{(u)}}\}$ as a function of $\tau (> \tau^{\ast})$.
		(b) The log-log plot of (a). 
		The thick black line segment shows the guide line, 
		whose slope is 0.5.}
\end{figure}

Next we focus on the eigenvalues of Jacobi matrix 
for the eight values of $(\bar{x}_{4}, \bar{y}_{4})$ 
when $\tau > \tau^{\ast}$.
It is confirmed that the eigenvalues of each fixed point in $\{\bar{\theta}_{4}^{\rm{(s)}}\}$ 
are the same, and the same is also true for $\{\bar{\theta}_{4}^{\rm{(u)}}\}$.
Figure \ref{fig:Lambda_st_unst}(a) shows the maximum eigenvalues 
$\lambda_{4}^{\rm{(s)}}$ for $\{\bar{\theta}_{4}^{\rm{(s)}}\}$ and 
$\lambda_{4}^{\rm{(u)}}$ for $\{\bar{\theta}_{4}^{\rm{(u)}}\}$ 
as a function of $\tau$.
Red asterisks and blue circles correspond 
to $\lambda_{4}^{\rm{(u)}}$ and $\lambda_{4}^{\rm{(s)}}$, respectively.
It is found that $\{\bar{\theta}_{4}^{\rm{(s)}}\}$ are stable 
and $\{\bar{\theta}_{4}^{\rm{(u)}}\}$ are unstable  
since $\lambda_{4}^{\rm{(s)}} < 1$ and 
$\lambda_{4}^{\rm{(u)}} > 1$ for all $\tau > \tau^{\ast}$.
Furthremore it is found that $\lambda_{4}^{\rm{(s)}}$ and $\lambda_{4}^{\rm{(u)}}$ 
tend to become 1 as $\tau$ goes to $\tau^{\ast}+0$.
Therefore, saddle-node bifurcation occurs at $\tau=\tau^{\ast}$.
Note that as an asymptotic property for $\tau$ dependence 
of $\lambda_{4}^{\rm{(s)}}$ and $\lambda_{4}^{\rm{(u)}}$, 
the following scaling relations are confimed 
as shown in Fig. \ref{fig:Lambda_st_unst}(b): 
$|\lambda_{4}^{\rm{(s)}}-1|, |\lambda_{4}^{\rm{(u)}}-1| \sim 
(\tau - \tau^{\ast})^{0.5}$.
From this bifurcation analysis, it is concluded that 
the ultradiscrete states in the limit cycle 
of the tropically discretized negative feedback model 
emerge due to phase lock by saddle-node bifurcation 
at $\tau = \tau^{\ast}$.

\ \\

\textit{Discussion}: 
(i) First we comment on an additional meaning 
of the tropical discretization.
One of the equations in eq.(\ref{eqn:1-2}) can be rewritten as 
\begin{equation}
	x_{n+1} = r_{n}(\tau) x_{n}, \;\;\;\;
	r_{n}(\tau) \equiv \frac{1+\tau \frac{f_{1}(x_n, y_n)}{x_n}}
				{1 + \tau \frac{g_{1}(x _n, y_n)}{x_n}}. 
	\label{eqn:1-2-mp}
\end{equation}
This equation for $x_{n}$ can be interpreted 
as a multiplicative time evolution from $x_{n}$ to $x_{n+1}$, 
where $r_{n}$ is the change ratio.
$f_{1}$ and $g_{1}$ in $r_{n}$ are considered 
as magnification and reduction factors, respectively. 
$\tau$ can be treated as a coefficient controlling 
the degree of magnification and reduction.
Therefore, the tropical discretization seems to be suitable 
for discrete modeling of multiplicative processes, 
which is in contrast to an additive discretization 
such as Euler difference method.
%

\begin{figure}[t!]
	\begin{center}
		\includegraphics[width=7.5cm]{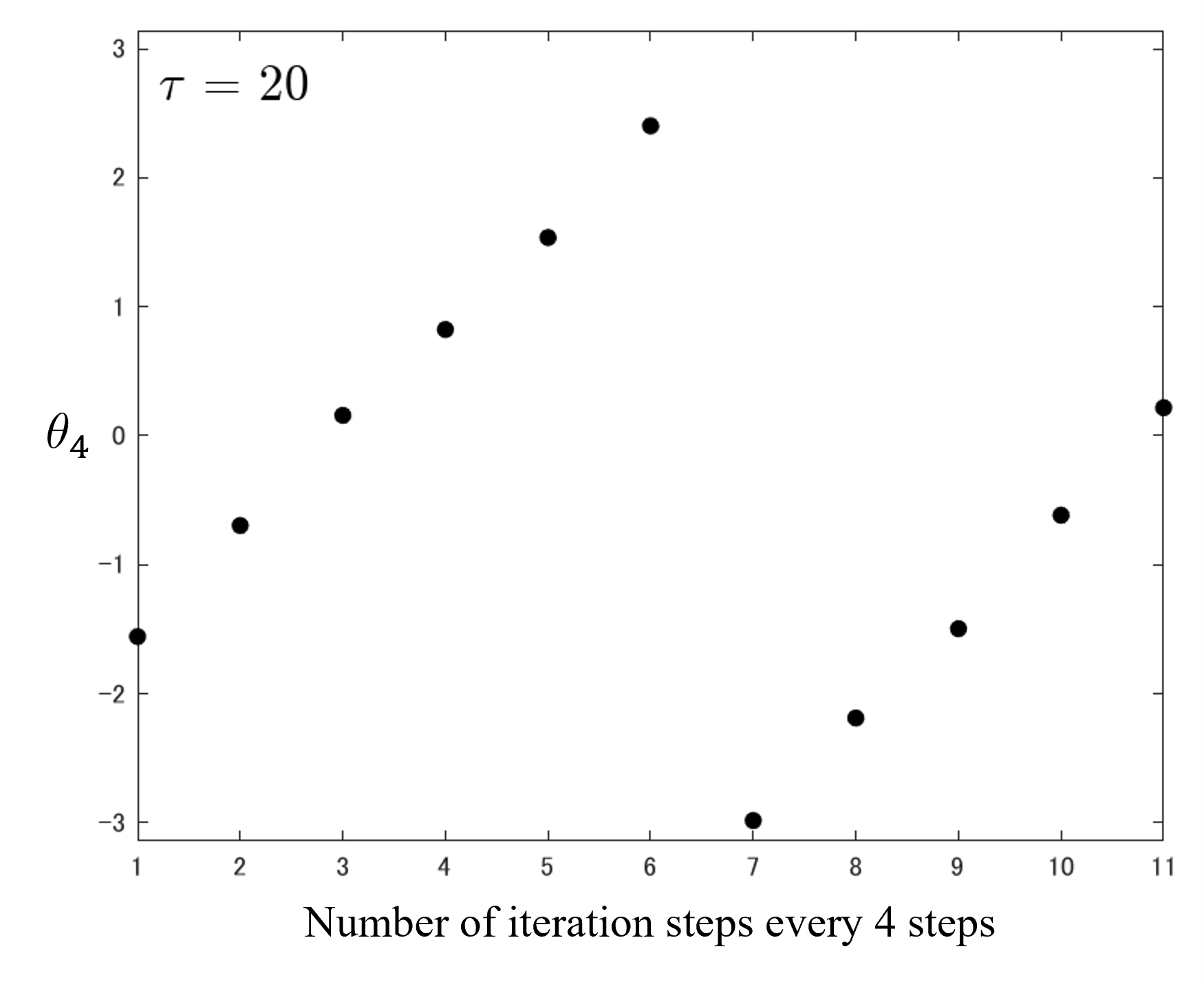}
		\includegraphics[width=7.5cm]{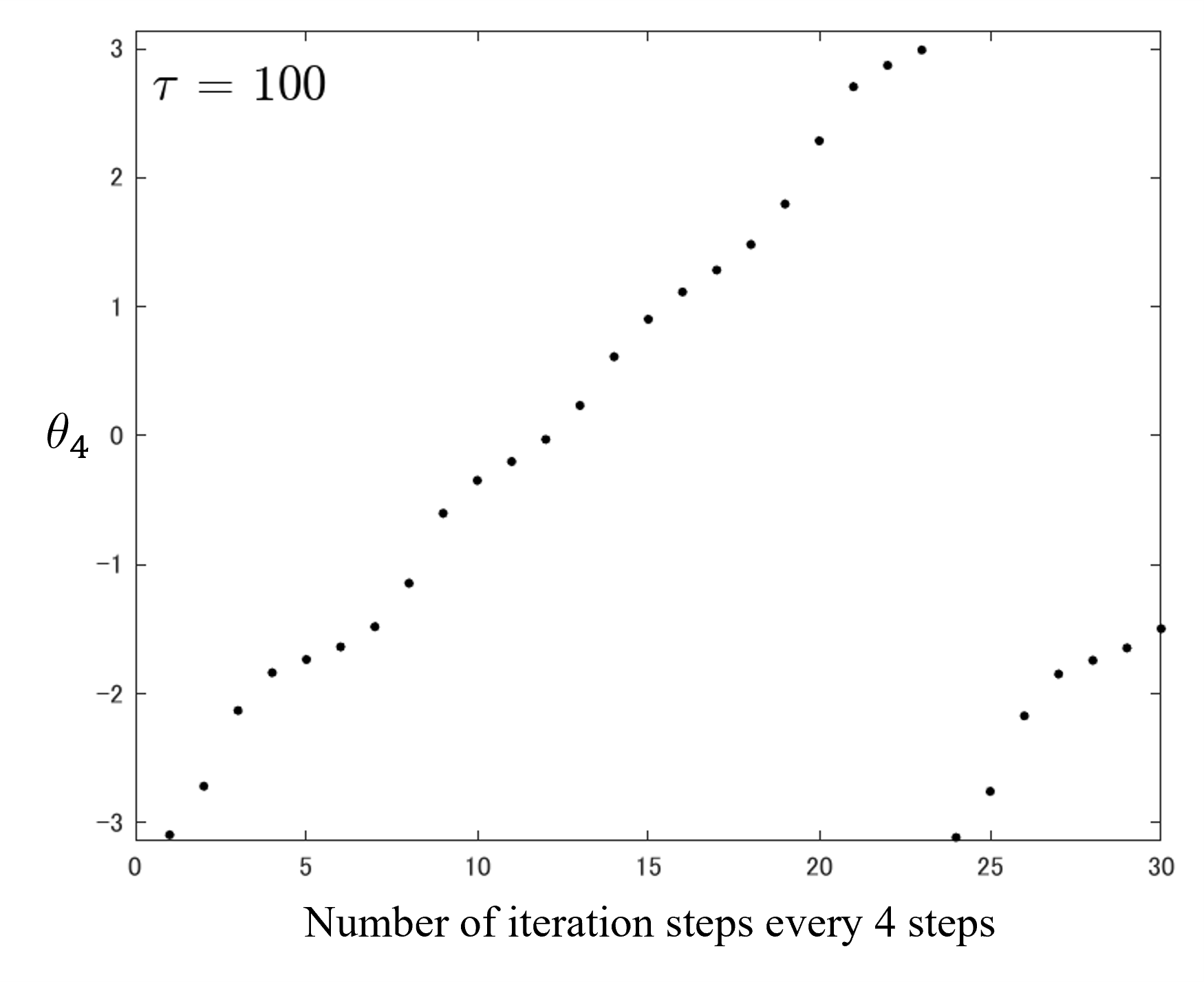}
	\end{center}
	\vspace{-8pt}
	\hspace{4.2cm}
	(a)
	\hspace{6.5cm}
	(b)	
	\begin{center}
		\includegraphics[width=7.5cm]{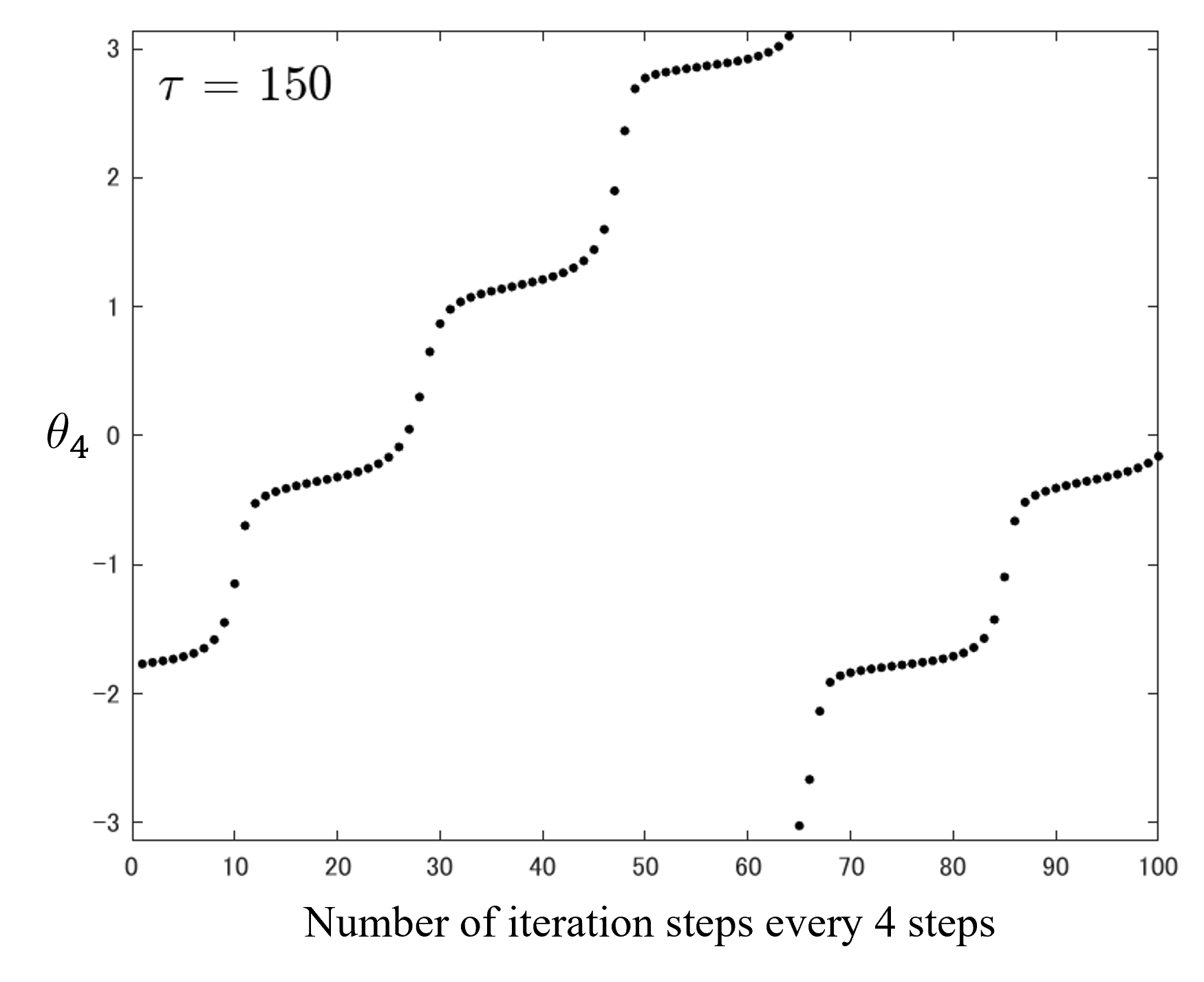}
		\includegraphics[width=7.5cm]{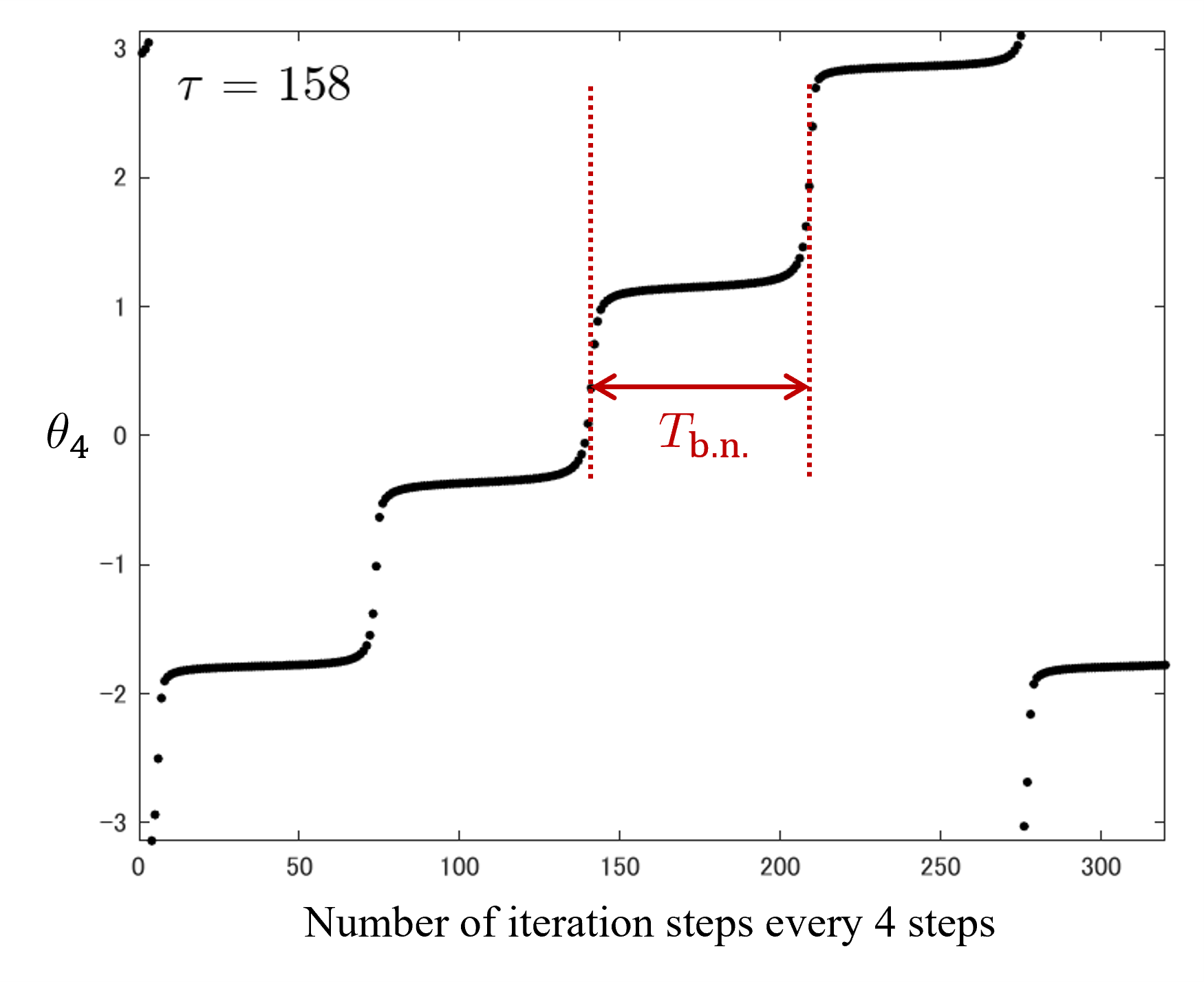}
	\end{center}
	\vspace{-8pt}
	\hspace{4.2cm}
	(c)
	\hspace{6.5cm}
	(d)	
	\begin{center}
		\includegraphics[width=7.5cm]{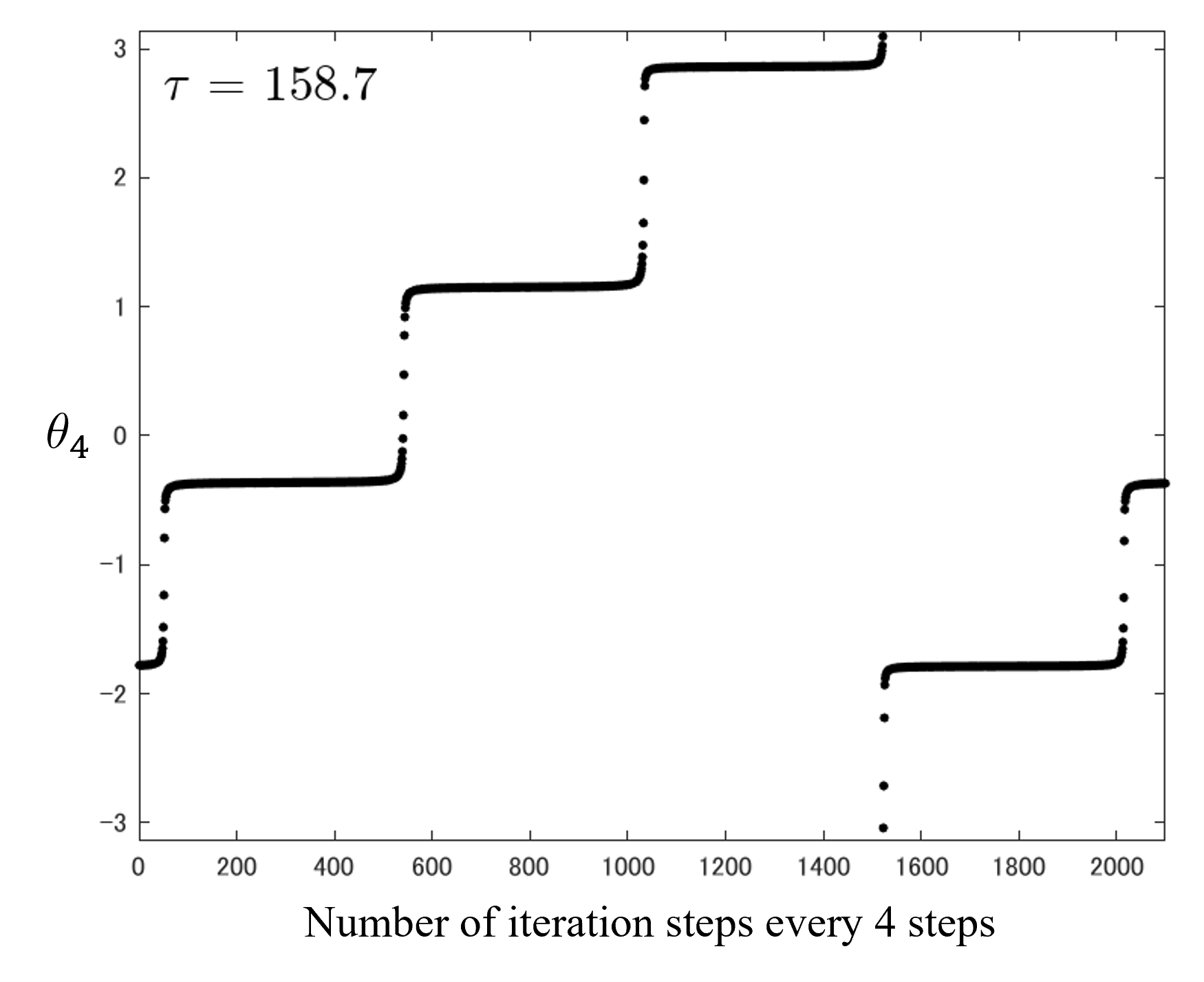}
		\includegraphics[width=7.5cm]{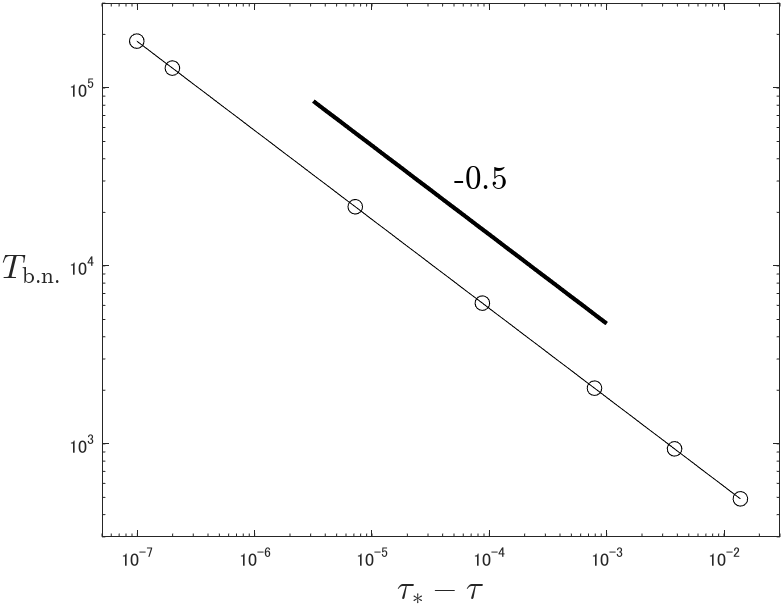}
	\end{center}
	\vspace{-8pt}
	\hspace{4.2cm}
	(e)
	\hspace{6.5cm}
	(f)	
	\caption{\label{fig:ts_phase} 
	  (a)-(e) Time evolutions of $\theta_{4n}$ with 
		five different values of $\tau$, where $b=10$ and $m=2$.
		(f) Scaling relation between $\tau^{\ast} - \tau$ and 
		$T_{\rm{b.n.}}$, which is defined as shown in (d).}
	\vspace{0.5cm}
	\end{figure}

(ii) When $\tau < \tau^{\ast}$ in eq.(\ref{eqn:tdnfb-x}), 
we found that $\theta_{4n}$ exhibits phase drift and bottleneck motion 
as shown in Fig.\ref{fig:ts_phase}(a)-(e).
Especially as $\tau$ approaches $\tau^{\ast}-0$, 
time to pass through the bottlenecks, 
which is defined as $T_{\rm{b.n.}}$ in Fig.\ref{fig:ts_phase}(d), 
becomes longer.
In fact as shown in Fig.\ref{fig:ts_phase}(f), 
$T_{\rm{b.n.}}$ is subject to the scaling relation:
$T_{\rm{b.n.}} \sim (\tau^{\ast} - \tau)^{-0.5}$.
Note that this scaling relation is well-known as 
the standard property in the vicinity of saddle-node bifurcation 
point\cite{Strogatz}.
Therefore, this slowing down in phase drift motion 
for $\tau \lesssim \tau^{\ast}$ also supports occurrence 
of saddle-node bifurcation at $\tau = \tau^{\ast}$.
%
%
%

(iii) The present results and conlusion are obtained from analysis 
of a specific model, the negative feedback model.
Similar tendency has been suggested 
for the Sel'kov model as well\cite{Ohmori2021,Yamazaki2021,Ohmori2022}.
Then the generality of this conclusion is expected, 
although further investigation is needed.
The introduction of phase in the analysis of limit cycles 
has long been a basic and usual method in nonlinear dynamics\cite{Kuramoto2003}.
The present study shows that the analysis on the basis of phase 
is also effective in the context of ultradiscretization, 
and it is interesting and important to review the study of Willox et al. \cite{Willox2007} 
from the viewpoint of phase dynamics.

\begin{figure}[b!]
	\begin{center}
		\includegraphics[width=7cm]{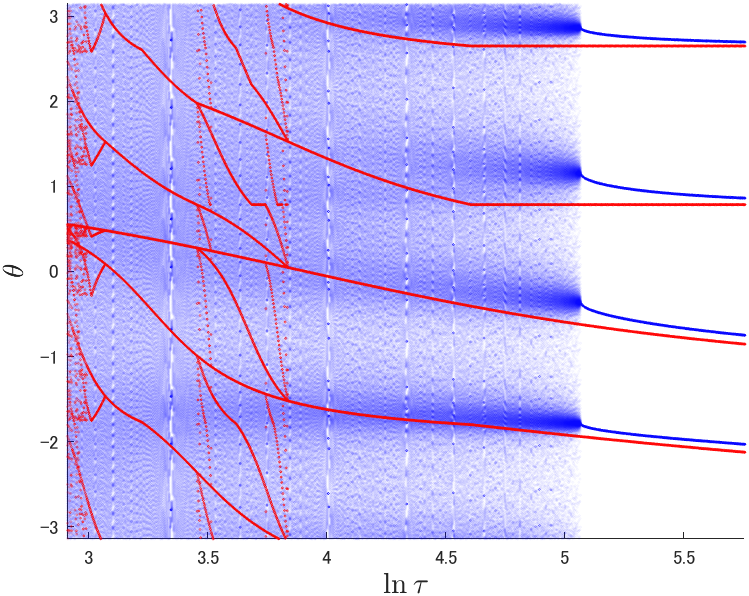}
		\includegraphics[width=7cm]{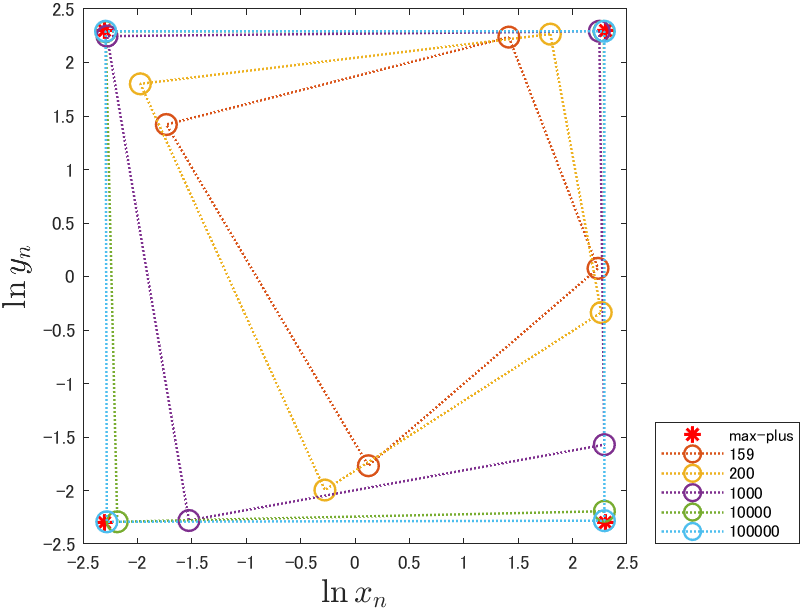}
	\end{center}
	\vspace{-8pt}
	\hspace{4.2cm}
	(a)
	\hspace{6.5cm}
	(b)	

	\caption{\label{fig:tropical_maxplus} 
	  (a) The red plots show $\{\Theta_{n}\}$ as a function of $\tau$ 
		obtained from eq.(\ref{eqn:2-3a}), where $m=2$ and $B=\ln b=\ln 10$.
		For each $\tau$, 500 points are plotted.
		The superimposed blue scatterd plot is the same as Fig.\ref{fig:lc_deltat}(a).
		(b) The open circles show limit cycle states with five different 
		values of $\tau > \tau^{\ast}$. 
		The asterisks show the limit cycle states 
		obtained from eq.(\ref{eqn:2-3a}).}
\end{figure}

(iv) Here we discuss relationship 
between limit cycle solutions obtained from eq.(\ref{eqn:tdnfb-x}) 
and those obtained from the max-plus equations, 
which are derived from eq.(\ref{eqn:tdnfb-x}).
Here in order to derive the max-plus equations from the discretized equations, 
we adopt the following replacement:
\begin{eqnarray}
	\displaystyle \log(e^{A_{1}}+e^{A_{2}}) \to \max(A_{1},A_{2}), 	
	\label{eqn:ud_tf}
\end{eqnarray}
where $A_{1}$ and $A_{2}$ are positive variables.
It is noted that eq.(\ref{eqn:ud_tf}) brings 
about piecewise linearization for the left hand side of eq.(\ref{eqn:ud_tf}), 
and is mathematically formulated as the limit identity  
by introducing an additional scaling parameter\cite{ultradiscrete}.
Introducing new variables $X_{n} \equiv \ln x_{n}$, 
$Y_{n} \equiv \ln y_{n}$, $B \equiv \ln b$, 
and $T \equiv \ln \tau$ in eq.(\ref{eqn:tdnfb-x}), 
and applying eq.(\ref{eqn:ud_tf}), 
we obtain the following max-plus equations.
\begin{equation}
	\begin{aligned}
		X_{n+1} & = \max(X_{n}, T+Y_{n}) - \max(0, T), \\
		Y_{n+1} & = \max(Y_{n}, T-\max(0, mX_{n})) - \max(0, T-B).
	\end{aligned}
	\label{eqn:2-3a}
\end{equation}
Performing numerical calculation of eq.(\ref{eqn:2-3a}), 
we confirmed that eq.(\ref{eqn:2-3a}) can possess cyclic solutions.
Then we introduce phase $\Theta_{n}$, which is defined as 
\begin{equation}
	\Theta_{n} = \arctan 
	\displaystyle \frac{Y_{n} - \ln \bar{y}}{X_{n} - \ln \bar{x}}.  
	\label{eqn:ud_theta}
\end{equation}
The red lines in Fig. \ref{fig:tropical_maxplus}(a) show 
scatter plot of $\{\Theta_{n}\}$ as a function of $\tau$, 
where the result shown in Fig.\ref{fig:lc_deltat}(a) is also 
superimposed on this figure.
For $\tau < \tau^{\ast}$, 
the phases obtained from eq.(\ref{eqn:tdnfb-x})
are broadly distributed and do not agree with the results 
obtained from eq.(\ref{eqn:2-3a}).
However for $\tau > \tau^{\ast}$, 
both limit cycles consist of four states, 
and we can confirm a quantitative trend toward agreement 
in the phase values as $\tau$ increases.
Actually, in Fig. \ref{fig:tropical_maxplus}(b), 
as $\tau$ increases, the states obtained by eq.(\ref{eqn:tdnfb-x})
approach the states obtained from eq.(\ref{eqn:2-3a}).
Note that for $B$ being finite, even when $\tau$ goes to infinity, 
the states by eq.(\ref{eqn:tdnfb-x}) 
do not exactly match the states by eq.(\ref{eqn:2-3a}) 
quantitatively due to piecewise linearization by eq.(\ref{eqn:ud_tf}).
Nevertheless, even when $\tau > \tau^{\ast}$ and $B$ is finite, 
it is meaningful to express the states 
by the max-plus equations  
in order to construct corresponding cellular automata.
(v) Here we consider the limit case of $\tau \to \infty$ ($T \to \infty$).
In this case, eq.(\ref{eqn:tdnfb-x}) and eq.(\ref{eqn:2-3a}) 
can be rewritten as 
\begin{equation}
	x_{n+1} = y_n \equiv \eta_{\infty}(x _n, y_n), \;\;\;\;
	y_{n+1} = \frac{b}{1+x_n^m} \equiv \xi_{\infty}(x _n, y_n), 
	\label{eqn:tdnfb-inf}
\end{equation}
\begin{equation}
	X_{n+1} = Y_{n}, \;\;\;\;
	Y_{n+1} = B - \max(0, mX_{n}).
	\label{eqn:2-3inf}
\end{equation} 
From our previous study\cite{Ohmori2023c}, 
it has been confirmed that there exist the two limit cycle solutions
in eq.(\ref{eqn:2-3inf}), $\mathcal{C}$ and $\mathcal{C}_{s}$, 
which consist of the following four states,  
$\displaystyle \mathcal{C}$: $(\pm B, \pm B)$ and 
$\displaystyle \mathcal{C}_s$: $\displaystyle \left(\frac{B}{3}, \pm B\right)$ 
and $\displaystyle \left(\pm B, \frac{B}{3}\right)$ for $m=2$.
The important point is that $\mathcal{C}_{s}$ 
is the unstable repulsive limit cycle.
Meanwhile, we focus on the fixed points for 4-th iterates of eq.(\ref{eqn:tdnfb-inf}), 
$(\bar{x}_{4\infty}, \bar{y}_{4\infty})$, 
which are obtained from 
$\bar{x}_{4\infty} = \eta^{4}_{\infty}(\bar{x}_{4\infty}, \bar{y}_{4\infty})$ 
and $\bar{y}_{4\infty} = \xi^{4}_{\infty}(\bar{x}_{4\infty}, \bar{y}_{4\infty})$. 
We numerically obtained the eight fixed points, 
from which $(\ln \bar{x}_{4\infty}, \ln \bar{y}_{4\infty})$ can be classified as 
$\mathcal{D}$: $(\pm \alpha, \pm \alpha)$
and $\mathcal{D}_{s}$: $(\beta, \pm \alpha)$ 
and $(\pm \alpha, \beta)$, 
where $\alpha$ and $\beta$ depend on the value of $B = \ln b$ 
as shown in Tbl.\ref{tbl:alpha_beta}.
It is found from Tbl.\ref{tbl:alpha_beta} that 
as $B$ increases, $\alpha$ and $\beta$ tend to $B$ and $B/3$, 
respectively.
Therefore, it is suggested that 
$\mathcal{D} \to \mathcal{C}$ and $\mathcal{D}_s \to \mathcal{C}_{s}$
when $B \to \infty$ and $T \to \infty$; 
this is consistent with arguments 
based on ultradiscrete limit formula\cite{ultradiscrete}.
We can say that $\mathcal{C}_{s}$ is the inheritance of $\mathcal{D}_{s}$ 
and the remnant of unstable fixed points 
caused by saddle-node bifurcation.
This result provides support for the fact 
that the dynamical structures of eq.(\ref{eqn:tdnfb-x}) 
are retained even in the simple max-plus form of eq.(\ref{eqn:2-3inf}).

\begin{table}[t!]
  \caption{Comparison of $\alpha$ and $\beta$  
	obtained from numerical calculation of eq.(\ref{eqn:tdnfb-inf}) 
	with the value of $B$.
	For $B - \alpha$ and $B/3 - \beta$, only the orders are presented.}
  \label{tbl:alpha_beta}
  \centering
  \begin{tabular}{cccccc}
    \hline
    $b$ & $B = \ln b$ & $\alpha$ & $\beta$ & $B - \alpha$ & $B/3 - \beta$ \\
    \hline \hline
    $10^1$    & 2.3025$\cdots$   &  2.2924$\cdots$ & 0.6931$\cdots$ & $10^{-2}$  & $10^{-2}$ \\
    $10^3$    & 6.9077$\cdots$   &  6.9077$\cdots$ & 2.2992$\cdots$ & $10^{-6}$  & $10^{-3}$ \\
    $10^5$    & 11.5129$\cdots$  & 11.5129$\cdots$ & 3.8374$\cdots$ & $10^{-10}$ & $10^{-4}$ \\
    $10^7$    & 16.1180$\cdots$  & 16.1180$\cdots$ & 5.3726$\cdots$ & $10^{-14}$ & $10^{-6}$ \\
    $10^9$    & 20.7232$\cdots$  & 20.7232$\cdots$ & 6.9077$\cdots$ & $10^{-18}$ & $10^{-7}$ \\
%
    \hline
  \end{tabular}
\end{table}
%

\textit{Summary and Conclusion}: 
We have investigated dynamical properties of the limit cycles 
for the tropically discretized negative feedback model.
Based on the bifurcation analysis with phase description, 
it is found that the limit cycle becomes ultradiscrete state with four states 
due to phase lock caused by saddle-node bifurcation at $\tau = \tau^{\ast}$, 
where $\tau$ behaves as the bifurcation parameter.
We have discussed relationship 
between limit cycle states obtained from tropically discretized 
and the max-plus models, and 
it is also found that the unstable limit cycle $\mathcal {C}_s$ 
in the max-plus model 
corresponds to the unstable fixed points $\mathcal {D}_s$ appearing 
by the saddle-node bifurcation in the tropically discretized model.
%
%

\noindent
{\bf Acknowledgement}\\
The authors are grateful to 
Prof. M. Murata, 
Prof. K. Matsuya, 
Prof. D. Takahashi, Prof. R. Willox, 
Prof. T. Yamamoto, and Prof. Emeritus A. Kitada 
for useful comments and encouragements. 
This work was supported by JSPS
KAKENHI Grant Numbers 22K13963 and 22K03442.


\begin{thebibliography}{9}
%
	\bibitem{Strogatz}
	S. H. Strogatz, \emph{Nonlinear Dynamics and Chaos} (Westview Press, U.S. 1994).
%
	\bibitem{Willox2007}
	R. Willox, A. Ramani, J. Satsuma, and B. Grammaticos,
	Physica A {\bf 385} 473 (2007).
	%
	\bibitem{Gibo2015}
	S. Gibo and H. Ito, 
	J. Theor. Biol., {\bf 378} 89 (2015).
	%
	\bibitem{Ohmori2016}
	S. Ohmori and Y. Yamazaki, J. Phys. Soc. Jpn. {\bf 85} 045001 (2016).
	%
	\bibitem{Ohmori2021}
	S. Ohmori and Y. Yamazaki, arXiv:2107.02435v1
	%
	\bibitem{Yamazaki2021}
	Y. Yamazaki and S. Ohmori, J. Phys. Soc. Jpn. {\bf 90} 103001 (2021).
	%
	\bibitem{Ohmori2022}
	S. Ohmori and Y. Yamazaki, 
	JSIAM Letters \textbf{14}, 127 (2022).
	%
	\bibitem{Ohmori2023c}
	S. Ohmori and Y. Yamazaki, 
	arXiv:2305.05908.
	%
		\bibitem{Isojima2022}
	S. Isojima and S. Suzuki, Nonlinearity.  {\bf 35} 1468 (2022).
	%
	\bibitem{Murata2013}
	M. Murata, J. Differ. Equations Appl. {\bf 19} 1008 (2013).
	%
	\bibitem{Carstea2006}
	A. S. Carstea, A. Ramani, J. Satsuma, R. Willox, and B. Grammaticos, 
	Physica A {\bf 364} 276 (2006).
	%
	\bibitem{Griffith1968}
	J. S. Griffith, 
	J. Theor. Biol. {\bf 20}, 202 (1968).
	%
	\bibitem{Selkov1968}
	E. E. Sel$^{\prime}$kov, Eur. J. Biochem. {\bf 4} 79 (1968).
	%
	\bibitem{Ohmori2023a}
	S. Ohmori and Y. Yamazaki, arXiv:2304.01573.
	%
	\bibitem{Kuramoto2003}
	Y. Kuramoto, 
	\textit{Chemical Oscillations, Waves, and Turbulence}
	(Dover Publications, New York, 2003).


	\bibitem{ultradiscrete}
		%
		For deriving max-plus description, 
		it is common to apply the ultradiscrete limit formula 
		expressed in the following equation 
		instead of eq.(\ref{eqn:ud_tf}) \cite{Tokihiro2004}.
		\[
			\displaystyle\lim_{\varepsilon  \to +0} \varepsilon  \log(e^{A_{1}/\varepsilon }+e^{A_{2}/\varepsilon }) = \max(A_{1}, A_{2}).
		\]
		%
		Here $\varepsilon$ is the additional scaling parameter, 
		and changing this value corresponds to controlling 
		the scale of viewing the systems. 
		%
		And taking this value to infinity corresponds 
		to the limit of zooming out, 
		which brings about piecewise linearization.
		%
		If we set $X_{n}^{\prime} \equiv \varepsilon \ln x_{n}$, 
		$Y_{n}^{\prime} \equiv \varepsilon \ln y_{n}$, 
		$B^{\prime} \equiv \varepsilon \ln b$, 
		and $T^{\prime} \equiv \varepsilon \ln \tau$ in eq.(\ref{eqn:tdnfb-x}), 
		and applying the above ultradiscrete limit formula, 
		we obtain the same max-plus equations as eq.(\ref{eqn:2-3a}).
		%
		In the main text, it can be said 
		that the case $\varepsilon=1$ is considered.
		%
	

	\bibitem{Tokihiro2004}
	T. Tokihiro, 
	\textit{Discrete Integrable Systems} 
	(edited by B. Grammaticos, T. Tamizhmani, and Y. Kosmann-Schwarzbach,
	Springer, Berlin, Heidelberg, 2004), pp. 383–424.



\end{thebibliography}
\end{document}